\documentclass[aps,prl,twocolumn,groupedaddress]{revtex4}

\usepackage{graphicx}% Include figure files

\begin{document}

%\preprint{}

\title{Contraction-free quantum state encoding by quantum tunneling in single molecules}


\author{Tomofumi Tada}
\email[]{tada.tomofumi.054@m.kyushu-u.ac.jp}
%\thanks{}
\affiliation{Kyushu University Platform of Inter/Transdisciplinary Energy Research, Kyushu University, 744 Motooka, Nishi-ku, Fukuoka 819-0395, Japan \\}
%\affiliation{Materials Research Center for Element Strategy (MCES), Tokyo institute of Technology, 4259 Nagatsuta, Midori-ku, Yokohama 226-8501, Japan \\}

\author{Masateru Taniguchi}
\email[]{taniguchi@sanken.osaka-u.ac.jp}
%\thanks{}
\affiliation{The Institute of Scientific and Industrial Research, Osaka University, 8-1 Mihogaoka, Ibaraki, Osaka 567-0047, Japan \\}


\date{\today}

\begin{abstract}
Quantum computing is a unique computational approach that promises tremendous performance that cannot be achieved by classical computers, although several problems must be resolved to realize a practical quantum computing system for easy use. Here, we propose a new system and theory for quantum computing that employs single molecule confinement between electrodes. The striking features of this system are (i) an individual molecule that exhibits quantum tunneling can be regarded as a sequence of quantum gates, (ii) the quantum tunneling can be encoded onto an array of quantum bits and observed without the contraction of superposition states, and (iii) quantum computing by quantum tunneling can be performed at room temperature. An adenine molecule is adopted as the single molecule between electrodes, and conductance data are encoded onto quantum states including entangled states, depending on the conductance values. As an application of the new quantum system, molecule identification based on quantum computing by quantum tunneling is demonstrated.
\end{abstract}

%\keywords{}

\maketitle


Quantum computing has been considered to open a new paradigm in computation, and quantum systems based on super-conductors, ion trapping, and impurities in solids have been developed as existing platforms for quantum computing \cite{ladd}. 
A quantum algorithm that provides quantum systems with tremendous performance as quantum computers has also been developed \cite{nielsen}. 
Although the road-map toward practical quantum computers is presented with many wonderful dreams, an extremely steep road toward the realization of quantum computers prevents the realization of a “quantum world”. There are fundamental difficulties in the realization of quantum computing: i) quantum systems that exhibit sufficient potential for quantum computing are limited to a few systems to date, ii) a very low temperature is required to preserve the quantum states of quantum bits (i.e., quantum information), and iii) the observation of quantum states, in principle, leads to the contraction of a superposition state such as $\left|0+1\right\rangle$  into a pure (i.e., non-superposition) state such as $\left| 0\right\rangle$ or $\left|1\right\rangle$. 
The first and second points would be disadvantageous in the development of practical quantum computers as physical systems, and the third point is an inherent problem related to observation in quantum mechanics; the contraction of quantum states by observation prohibits acquisition of the quantum-mechanical results in a superposition state (e.g.,  $\left|0+1\right\rangle$). 
Therefore, quantum computations and observations must be executed with a quantum computer in a repeated manner to obtain a definitive result.
Note that even when quantum computing with a repeated manner is required, the advantage of quantum computers is still obvious, especially in specific problems. However, let us consider what is expected if a single quantum execution is sufficient to obtain a definitive result. If an observation is accomplished without the contraction of superposition states, then single quantum execution is realized, and the advantage of quantum computing would be obvious for a wide range of problems. Here we propose a completely new system and theory to resolve the difficulties in quantum computing by using “existing” single molecule confinement (SMC) between electrodes \cite{taniguchi1} that works at room temperature. The striking features of this quantum system and theory are (i) an individual molecule in SMC can be regarded as a specific quantum computing system defined by quantum-gates derived from the molecular orbital rule for quantum tunneling \cite{tada1}, (ii) quantum-mechanical results including superposition states using SMC can be observed as is, and (iii) the system works at room temperature. The paper is organized as follows; I) a brief introduction of SMC, II) contraction-free quantum state observation, III) contraction-free quantum state encoding,  IV) the theoretical foundation for the contraction-free quantum state observation, and V) conclusion.

\section{Single Molecule confinement}
\begin{figure}
\includegraphics[scale=0.5]{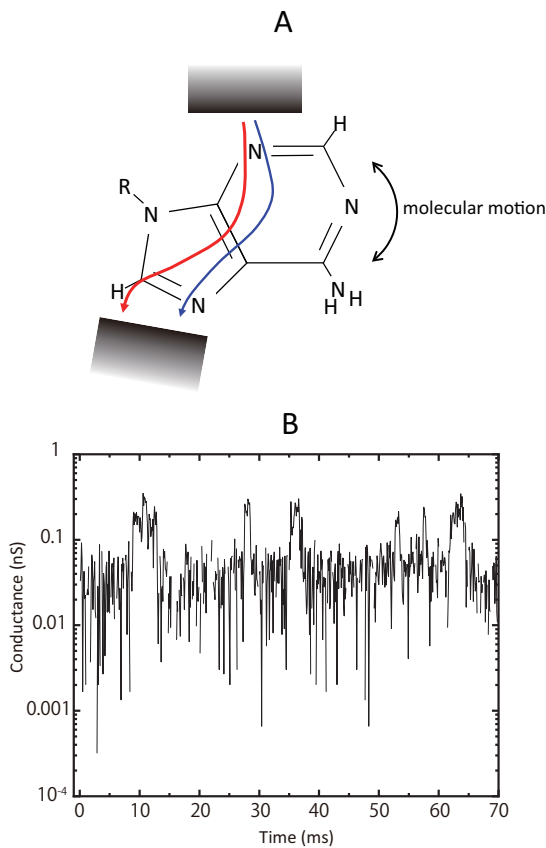} 
\caption{\label{1DNMRQ} Single molecule confinement and observed conductance data. (A) Schematic diagram of single molecule (adenine) confinement between nanogap electrodes. (B) Observed conductance of single adenine molecule confinement at room temperature. }
\end{figure}

\begin{figure}
\includegraphics[scale=0.4]{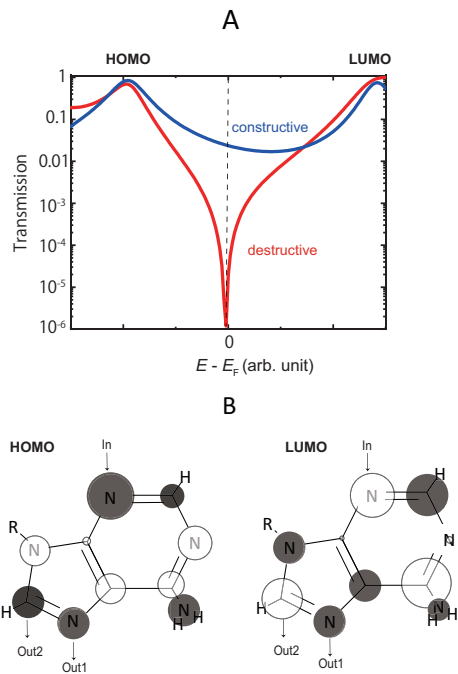} 
\caption{\label{SNuSF} Quantum tunneling of adenine single molecule confinement. (A) Calculated transmission function of single adenine confinement with constructive (blue) and destructive (red) tunneling pathways. (B) Highest occupied molecular orbital (HOMO) and lowest unoccupied molecular orbital (LUMO) of the adenine single molecule, which are useful to predict the constructive and destructive tunneling pathways. }
\end{figure}

Figure 1 shows a schematic of SMC and an example of the observed quantity, i.e., the conductance of SMC \cite{taniguchi1}. 
When the size of the system sandwiched between the electrodes is macroscopically large, the current is simply characterized by Ohm's law; however, the situation is changed when the sandwiched system is quite a small object, such as mesosized rings or nanosized molecules. In such a small ring, a superposition state in terms of current pathways is realized; the quantum interference of current pathways is known as the Aharonov-Bohm effect  \cite{AB}. 
In molecular systems, on the other hand, the situation becomes slightly more complicated, and a superposition state of molecular {\it tunneling orbitals} (frontier orbitals coupled with electrodes) is an appropriate scenario to understand the tunneling current through SMC, which has been recognized as the quantum interference in single molecules  \cite{tada1,yoshizawa1,taniguchi2,guedon,li}. That is, in small-sized systems, tunneling states are inherently quantum states including superposition states, which can be a unit for quantum information, a quantum bit.  

In tunneling current measurements of SMC, a small electrical bias is typically applied to single molecules between electrodes, and the current or conductance is analyzed, depending on the purpose. The conductance is slightly more convenient than the current in the present purpose, and we thus consider the conductance data shown in Fig. 1B, which is measured conductance of a single DNA-base molecule, adenine, with a SMC system \cite{taniguchi1}.

Now, we briefly introduce a molecular orbital rule for quantum tunneling to understand the large conductance variation over a few orders of magnitude. In what follows, we simply use the term “tunneling orbital rule” to indicate the molecular orbital rule for quantum tunneling \cite{tada1}.
According to the tunneling orbital rule, the conductance is highly correlated to the in-coming and out-going positions (i.e., sites in the molecule) of tunneling electrons. The key conclusion derived from the tunneling orbital rule is that there are construction/destruction positions to enhance/decrease the conductance (Fig. 2A), and the positions can be simply predicted with the molecular orbitals (MOs) of the confined molecule.
Figure 2B shows the $\pi$ highest occupied MO (HOMO) and lowest unoccupied MO (LUMO) of adenine. 
 The different colors (black/white) correspond to the difference of the sign (positive/negative) of the MO coefficients of the HOMO and LUMO. Using the tunneling orbital rule, we know that constructive/destructive interference in quantum tunneling occurs when the sign of the multiplied HOMO coefficient at the two positions contacting with the electrodes (i.e., in-coming and out-going positions) is different/same to that of the multiplied LUMO coefficient at the same two positions. In Fig. 2B, the pair of “In” and “Out1” positions corresponds to the constructive interference, and the pair of “In” and “Out2” corresponds to the destructive interference.
The tunneling orbital rule is straightforwardly derived from the unperturbed Green's function of the molecule, 
$\sum_{i} \frac{|\phi_{i} ><|\phi_{i}|}{E - E_{i}}$, where $E$, $E_{i}$, and $\phi_{i}$ are the energy of tunneling electrons, the $i$-th MO energy, and the $i$-th wavefunction of the molecule, respectively. 
Calculated transmission functions $T(E)$ of tunneling particles based on the Green's function are shown in Fig. 2A  for both constructive and destructive interference. 
The expression of the Green's function for the molecule indicates the two peaks of $T(E)$ are originated from the HOMO and LUMO.
According to the Landauer model \cite{landauer}, when a small bias voltage is applied to the SMC, the conductance of the SMC is proportional to the transmission function at the Fermi level of the electrodes. The Fermi level of the electrodes is typically located at around the mid-position between the HOMO and LUMO, and thereby the conductance of the destructive interference can be smaller than that of the constructive interference by a few or more orders of magnitude. The large variation of conductance for the SMCs shown in Fig. 1B is thus recognized as the result of constructive and destructive interference. The theoretical details for the molecular Hamiltonian and conductance calculations are explained in Appendix.
%Even when the Fermi level is shifted from the mid-position, we can expect a large difference of conductance between the constructive/destructive interference. 

It should be noted that the SMC described in this study is different from so-called single molecular junctions, in which both sides of the contact between a single molecule and electrodes are established with covalent chemical bonds. In single molecular junctions, sulfur-gold bonds are typically adopted to make strong contacts. Although the conductance of single molecular junctions can fluctuate by molecular vibrations, the magnitude of the fluctuations cannot be so large as to change the order of conductance, because molecular vibrations result in a small fluctuation of conductance. In fact, such a small fluctuation of conductance can also be confirmed in Fig. 1B.
%In this sense, we don't exclude a molecular confinement situation as SMC where one side is tightly bonded with an electrode (e.g., via S-Au) but another side is not binded by any covalent bonds.

\section{Contraction-free quantum state observation}

Let us consider again the large variation of conductance shown in Fig. 1B. We can recognize that the highest values of conductance (i.e., ca. 0.1 nS or more in Fig. 1B) originate from the constructive interference in tunneling and the lowest values (i.e., ca. 0.01 nS or less in Fig. 1B) originate from destructive interference in tunneling, because the difference in conductance between constructive and destructive interference can be larger than a few orders of magnitude, in principle. Now, we consider the intermediate values of conductance. Let us begin with a “classical” picture. A confined molecule can show a large structural deformation together with molecular vibrations; therefore, such a large deformation could be one candidate to explain the intermediate conductance. However, in such a deformed case, several variations of deformation could be expected, and thus a wide range of conductance values must be observed as intermediate conductance. The intermediate conductance shown in Fig. 1B is rather a definitive value, and thus deformation of molecular structure is insufficient to explain the intermediate conductance. The adenine molecule is a planar molecule, and thus a large deformation is not expected. Although we have considered the variations of contact positions as another reason, the two pairs for the contact (In-Out1 and In-Out2 shown in Fig. 2B) were determined as energetically possible, even though other pairs are energetically impossible, which indicates the variations of contact positions are also insufficient to explain the intermediate conductance. Please refer to the calculated total energies listed in Table. A1 (Appendix). 

Now, we move on to a “quantum” picture. In the tunneling orbital rule, it was assumed that a single site is selected as an “In” or “Out” position; however, this assumption means that a single quantum (i.e., an electron) must always select a single site to enter/escape into/from a single molecule. This is a somewhat too simplified assumption in quantum tunneling, i.e., in quantum mechanics, the superposition of tunneling pathways between In-Out1 and In-Out2 in Fig. 3A will be easily realized. The calculated transmission function that takes the superposition of the two pathways into account is shown as the black solid line in Fig. 3A; the blue/red lines correspond to the constructive/destructive interference (see Appendix for the computational conditions adopted in the transmission calculations).
Unexpectedly, the resultant conductance in the pathway superposition shows almost the same conductance as the constructive interference, i.e., the conductance cannot be apparently smaller conductance in the superposition of tunneling pathways. The next quantum point to be considered is the superposition of molecular configurations with respect to the electrodes because nuclei are also quantum mechanical particles that exhibit tunneling between stable positions. Fig. 3B shows the superposition of the two configurations that correspond to constructive and destructive tunneling. In this situation, the calculated conductance (black line in Fig. 3B) is an average between the constructive and destructive tunneling, and the value of the conductance is almost half that of the constructive interference. The constructive and destructive tunneling and the configuration superposition between them result in conductance of 0.132, 0.001, and 0.066 nS, respectively, as shown in Fig. 3B. According to the observation, it was concluded that the high and low conductance can be respectively assigned to the constructive and destructive interference, and that the intermediate conductance is the superposition state between the constructive and destructive molecular configurations for tunneling. The theoretical foundation for the tunneling processes including a superposition state between molecular configurations is explained in the theoretical part.

\begin{figure}
\includegraphics[scale=0.33]{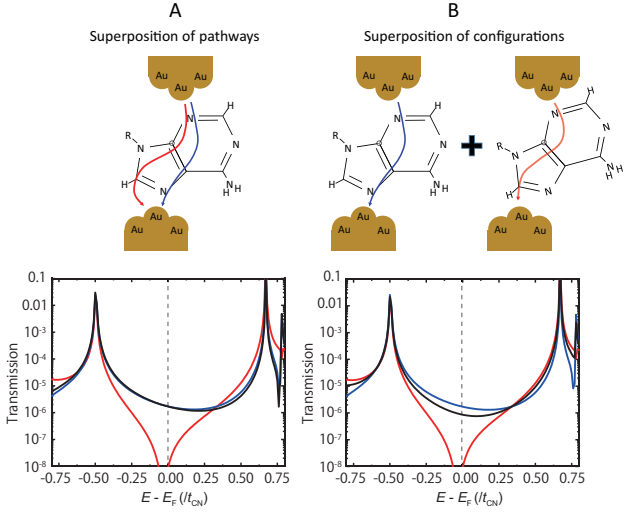} 
\caption{\label{transm} Superposition of tunneling pathways vs. superposition of molecular configurations. Calculated transmission functions for the superposition of (A) tunneling pathways and (B) molecular configurations. The blue/red lines correspond to constructive/destructive transmission, and the black line corresponds to the superposition state. The unit of energy ($t_{\rm CN}$) corresponds to the transfer energy between 2p$_{\pi}$(C) and  2p$_{\pi}$(N), 4.07 eV.}
\end{figure}

\section{Contraction-free quantum state encoding}

Next we explain how the quantum states including superposition states can be encoded onto a bit array. Before the encoding of a superposition state, the constructive (i.e., high conductance) and destructive (i.e., low conductance) tunneling should be encoded as pure states in terms of $\left|0\right\rangle$ and $\left|1\right\rangle$. In accordance with the tunneling processes shown in Figs. 1A and 2B, a bit array composed of three quantum bits (Qbits) is the minimum model; Qbits 1, 2, and 3 correspond to the sites of In, Out1, and Out2, respectively, shown in Fig. 2B. Here, we consider the initial state as $\left|100\right\rangle$, which indicates a tunneling electron injected at the In site, i.e., the states $\left|1\right\rangle$ and $\left|0\right\rangle$ respectively correspond to electrons passing through or not passing through. After the tunneling, we can expect several patterns as a bit array. For example, the tunneling process from In to Out1 can be encoded as $\left|110\right\rangle$, and that from In to Out2 as $\left|101\right\rangle$. In addition, the superposition state between $\left|110\right\rangle$ and $\left|101\right\rangle$, i.e., $\left|110\right\rangle + \left|101\right\rangle$,  a quantum entangled state, is required.
%If we can produce such quantum states including the superposition state by a sequence of quantum gates, the sequence of quantum gates corresponds to the quantum gates represented by the tunneling process of the molecule shown in Fig. 1A. 
A sequence of quantum gates that produce the pure states as $\left|110\right\rangle$ and $\left|101\right\rangle$ and the superposition state as $\left|110\right\rangle + \left|101\right\rangle$ is constructed using the unitary and controlled-NOT gates, as shown in Fig. 4A. Each tunneling process (i.e., constructive, destructive, and the superposition) is clearly confirmed, as demonstrated in Fig. 4B, i.e., the quantum tunneling process and conductance measurements with SMC correspond to the operation and read-out in a quantum computer, respectively, and a superposition state is observed in a contraction-free manner as the intermediate conductance. 
%Please see Supplementary materials for an instruction of how to construct the quantum gates for SMC in a well defined manner. Supplementary materials also include examples for other quantum gates corresponding to other tunneling patterns.  

\begin{figure}
\includegraphics[scale=0.5]{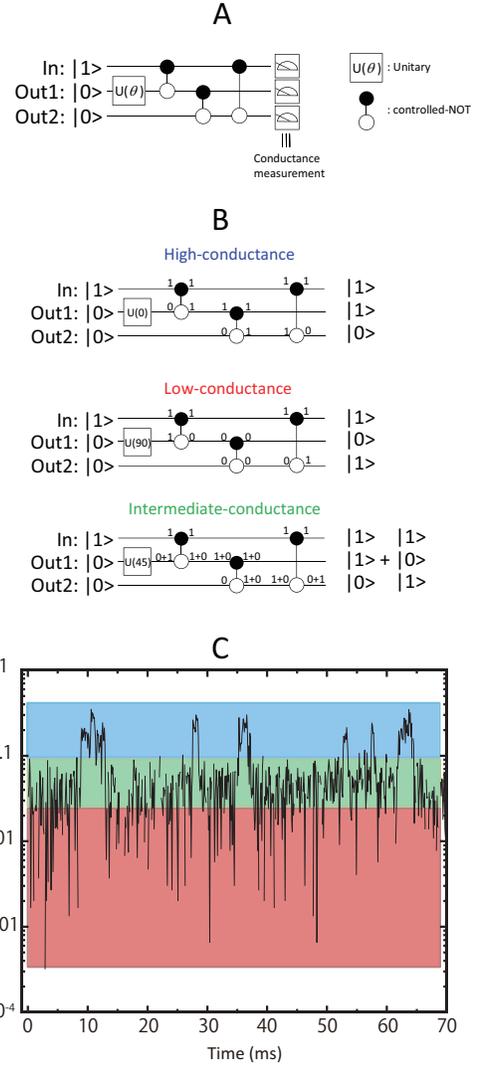} 
\caption{\label{transm} Contraction-free quantum state encoding. (A) Sequence of quantum gates constituted by unitary and controlled-NOT gates that correspond to single adenine molecule confinement. (B) Quantum computing processes that correspond to the constructive, destructive, and the superposition tunneling states through adenine. (C) Conductance data with assignments of the constructive (blue; $\left|110\right\rangle$), destructive (red; $\left|101\right\rangle$), and entangled superposition (green; $\left|110\right\rangle$ + $\left|101\right\rangle$) states.}
\end{figure}

A unitary gate defined with the rotation angle $\theta$ is included in the sequence of quantum gates shown in Fig. 4A. As expected, any weights (real numbers) for the constructive and destructive configurations as the superposition state can be realized. In other words, the distribution of $\theta$ represents a signature of the quantum mechanical feature of the confined single molecule, and this is a “quantum kernel” for a quick identification of single molecules.

The distribution of $\theta$ can be obtained for any single molecules using a corresponding sequence of quantum gates and conductance data. Thus, molecular identification using the sequence of quantum gates in a back-flow mode with $-\theta$ is accomplished using an existing quantum-gate quantum computer. Let us examine a simple case where the sequence of quantum gates for molecule A (known) is the same as that shown in Fig. 4A, and $\theta$ takes the values of $0^\circ$, $45^\circ$, and $90^\circ$ with the same probabilities (i.e., 33.3\%). Now let us consider the conductance data of molecule X (unknown). When the conductance data of molecule X are explicitly categorized into three levels of conductance with the same probabilities, and the intermediate conductance is half of the highest conductance, the back-flow mode of quantum computations always leads to a result of $\left|100\right\rangle$, i.e., the identification of molecule X as molecule A is 100\%. On the other hand, when the conductance data are categorized into three levels but have different probabilities than each other (or the intermediate conductance is not half of the highest conductance), the identification percentage as molecule A will not be 100\% but less than 100\%. If the conductance data of molecule X are scattered (i.e., not three levels), then the back-flow mode of quantum computations will not result in $\left|100\right\rangle$ in many quantum executions, i.e., a very small identification percentage as molecule A will be obtained. The back-flow mode can be performed with a quantum-gate quantum computer; therefore, the identification percentage can be quickly obtained. The above example is a very simple one, and much more detailed demonstrations of molecular identification using our quantum state encoding will appear soon in the future. 

\section{Theoretical foundation for the contraction-free quantum state observation}
 In this section, we provide the theoretical foundation for the contraction-free observation of quantum states. We apply the following two-step explanation: i) we explain how a superposition state composed of molecular states such as HOMO and LUMO can be preserved (i.e., observed) during tunneling current measurements, and ii) we extend the formulation for a superposition state between molecular configurations, and explain how the transmission function is modified in the configuration superposition.  The explanations in Step (i) basically follow the theoretical foundation for tunneling current given by Emberly and Kirczenow.\cite{emberly} Let us start from Step (i). The normalized wavefunction ${\rm \bf \Psi}$ of the system shown in Fig. A1C (Appendix) can be represented simply as

\begin{equation}
\left| {\rm \bf \Psi} \right\rangle =\sum^{-1}_{n=-\infty} \psi_{n} \left| n \right\rangle + \sum^{\infty}_{n=1} \psi_{n} \left| n \right\rangle  + \sum_{j} c_{j} \left| \phi_{j} \right\rangle , 
\label{eq7}
\end{equation}
where  $\left| n \right\rangle (n=-\infty,\cdots,-1,1\cdots, \infty)$ is an orthonormal basis for electrodes, and $\left| \phi_{j} \right\rangle (j=1,2, \cdots)$ is the molecular orbital of the sandwiched molecule. The expansion coefficients, $\psi_n$ and $c_j$, are determined depending on the interactions $W$ between the sandwiched molecule and electrodes. Lippmann-Schwinger (LS) equation is useful to understand how the wavefunction  ${\rm \bf \Psi}$ is modified by the interactions. Let us consider that electrons are injected from the left electrode and transmitted to the right-hand side. If there is no interactions between the left electrode and the rest (i.e., the molecule and right-electrode), the electron is represented with the eigenstate of the left-electrode, $\left| {\rm \bf \Phi}_{0} \right\rangle (=\sum^{-1}_{n=-\infty}  (\phi_{0})_{n} \left| n \right\rangle )$ , and when the interactions happen, the wavefunction is represented with LS equation as
%We first consider a specific case where there is no interactions between the single molecule and electrodes (i.e., the term $w$ in Fig. S1 is zero). Considering that electrons are injected from the left-electrode and that tunneling electrons are observed at the right-electrode, the observable can be represented as $\left\langle 1 |{\rm \bf \Psi} \right\rangle$, which is identical to $\psi_{1} \left\langle 1 |1\right\rangle$ (= $\psi_{1}$). This is a contraction as a usual case, that is, a contraction from$ \left| {\rm \bf \Psi} \right\rangle $ to $\left| 1 \right\rangle$. 
\begin{equation}
\left| {\rm \bf \Psi} \right\rangle =\left| {\rm \bf \Phi}_{0} \right\rangle + G_{0}W\left| {\rm \bf \Psi} \right\rangle ,
\label{eq8}
\end{equation}
where $ G_{0}$ is the Green's function for the decoupled system of the left- and right-electrodes and the molecule. Using Eqs.~(\ref{eq7}) and (\ref{eq8}), we obtain 
\begin{eqnarray}
\left| {\rm \bf \Psi} \right\rangle &=& \sum^{-1}_{n=-\infty} (\phi_{0})_{n} \left| n \right\rangle 
 + ( \sum^{-1}_{n,m=-\infty} (G^{L}_{0})_{n,m}  \left| n \right\rangle \left\langle m \right|     \nonumber   \\
 &+& \sum^{\infty}_{n,m=1} (G^{R}_{0})_{n,m}  \left| n \right\rangle \left\langle m \right|  
+ \sum_{j} (G^{M}_{0})_j  \left| \phi_j \right\rangle \left\langle \phi_j \right| ~)       \nonumber   \\
&\times& W \times(\sum^{-1}_{n=-\infty} \psi_{n} \left| n \right\rangle + \sum^{\infty}_{n=1} \psi_{n} \left| n \right\rangle  + \sum_{j} c_{j} \left| \phi_{j} \right\rangle~) . \nonumber   \\
& & 
\label{eq9}
\end{eqnarray}
When the electron is observed at the apex of the left-electrode, the wavefunction ${\rm \bf \Psi}$ is projected onto the base  $\left| -1 \right\rangle$ as
\begin{equation}
\left\langle -1 |{\rm \bf \Psi} \right\rangle = (\phi_{0})_{-1} + (G^{L}_{0})_{-1,-1} (\sum_{j} c_j \left\langle -1| W | \phi_j \right\rangle ) .
\label{eq10}
\end{equation}
Note that we used in Eq.~(\ref{eq10}) the fact that the integrals related to $W$ can be non-zero only for $\left\langle -1| W | \phi_j \right\rangle$ and  $\left\langle 1| W | \phi_j \right\rangle$ because $W$ is the interactions between the left/right-electrode and molecule. 
When there is no interactions between the left-electrode and the molecule, we obtain $\left\langle -1 |{\rm \bf \Psi} \right\rangle = (\phi_{0})_{-1} $, which is a standard conraction from $ \left| {\rm \bf \Psi} \right\rangle $ to $\left| -1 \right\rangle$. 

In the similar manner, we can consider the case where the electron is observed at the apex of the right-electrode (i.e., transmitted) as, 
\begin{equation}
\left\langle 1 |{\rm \bf \Psi} \right\rangle = (G^{R}_{0})_{1,1} (\sum_{j} c_j \left\langle 1| W | \phi_j \right\rangle )
\label{eq11}
\end{equation}
Transmission function is calculated as $\left|\left\langle 1 |{\rm \bf \Psi} \right\rangle\right|^2 $.  If there is no interactions between the right-electrode and the molecule, Eq.~(\ref{eq11}) reads $\left\langle 1 |{\rm \bf \Psi} \right\rangle = 0 $, which indicates the zero tunneling of electrons from the left-electrode to the right-electrode. 

Eq.~(\ref{eq11}) clearly tells us an important conclusion that the superposition of molecular eigenstates can be preserved (i.e., no-contraction) even after the observation of tunneling electron by the right-electrode. The coefficients ${c_j}$ can be represented with the molecular Green's function  $G^{M}_{0}$ \cite{emberly}, and the expansion of molecular orbitals in terms of atomic orbitals and HOMO-LUMO approximation in the Green's function lead to the molecular orbital rule for tunneling \cite{tada1}, where the superposition of HOMO and LUMO results in the constructive and destructive quantum interferences, as it has been experimentally confirmed \cite{li}. 

Now, we are ready to move on to Step ii). Here we consider two configurations, A and B, of a molecule with respect to the electrodes. When the molecule takes a single configuration (i.e., A or B), the normalized wavefunction of configuration A is written as 
\begin{equation}
\left| {\rm \bf \Psi}^A \right\rangle =\sum^{-1}_{n=-\infty} \psi_{n} \left| n \right\rangle + \sum^{\infty}_{n=1} \psi_{n} \left| n \right\rangle  + \sum_{j} c^A_{j} \left| \phi^A_{j} \right\rangle , 
\label{eq12}
\end{equation}
and that of configuration B is
\begin{equation}
\left| {\rm \bf \Psi}^B \right\rangle =\sum^{-1}_{n=-\infty} \psi_{n} \left| n \right\rangle + \sum^{\infty}_{n=1} \psi_{n} \left| n \right\rangle  + \sum_{j} c^B_{j} \left| \phi^B_{j} \right\rangle .
\label{eq13}
\end{equation}

Assuming that the amount of charge transfer between the molecule and electrodes is the same in the two configurations (in fact, this is a reasonable condition for SMC), the relation of $\sum_j \left| c^A_j \right|^2 = \sum_j \left| c^B_j \right|^2 $ is satisfied, and thereby the normalized wavefunction for the configuration superposition between A and B can be written as
\begin{eqnarray}
\left| {\rm \bf \Psi}^{A+B} \right\rangle & = & \sum^{-1}_{n=-\infty} \psi_{n} \left| n \right\rangle + \sum^{\infty}_{n=1} \psi_{n} \left| n \right\rangle  + \frac{1}{\sqrt 2}\sum_{j} c^A_{j} \left| \phi^A_{j} \right\rangle    \nonumber   \\
& +  & \frac{1}{\sqrt 2}\sum_{j} c^B_{j} \left| \phi^B_{j} \right\rangle,
\label{eq14}
\end{eqnarray}
where we adopted the relation of $\left\langle \phi^N_j |\phi^M_k \right\rangle = \delta_{NM} \delta_{jk}$.
Using Eqs.~(\ref{eq14}) and (\ref{eq8}), LS equation, the wavefunction with the configuration superposition is rewritten as 
\begin{eqnarray}
\left| {\rm \bf \Psi}^{A+B} \right\rangle &=& \sum^{-1}_{n=-\infty} (\phi_{0})_{n} \left| n \right\rangle 
 + ( \sum^{-1}_{n,m=-\infty} (G^{L}_{0})_{n,m}  \left| n \right\rangle \left\langle m \right|     \nonumber   \\
&+& \sum^{\infty}_{n,m=1} (G^{R}_{0})_{n,m}  \left| n \right\rangle \left\langle m \right|   \nonumber   \\
&+& \sum_{M=A,B} \sum_{j} (G^{M}_{0})_j  \left| \phi^M_j \right\rangle \left\langle \phi^M_j \right| ~)       \nonumber   \\
&\times& W \times ( \sum^{-1}_{n=-\infty} \psi_{n} \left| n \right\rangle + \sum^{\infty}_{n=1} \psi_{n} \left| n \right\rangle \nonumber   \\
& +& \frac{1}{\sqrt 2}\sum_{j} c^A_{j} \left| \phi^A_{j} \right\rangle + \frac{1}{\sqrt 2}\sum_{j} c^B_{j} \left| \phi^B_{j} \right\rangle ) .\nonumber   \\
& &
\label{eq15}
\end{eqnarray}
Therefore the observation of the tunneling electron at the right electrode reads 
\begin{eqnarray}
\left\langle 1 |{\rm \bf \Psi}^{A+B} \right\rangle &=&  \frac{1}{\sqrt 2} (G^{R}_{0})_{1,1}   \nonumber   \\
& \times& ( \sum_{j} c^A_j \left\langle 1| W | \phi^A_j \right\rangle + \sum_{j} c^B_j \left\langle 1| W | \phi^B_j \right\rangle ),  \nonumber   \\
& &
\label{eq16}
\end{eqnarray}
and we obtained the conclusion that the measurements of tunneling electron with an electrode does not break the superposition state between different molecular configurations, as well as that between molecular eigenstates in each configuration. This is the theoretical foundation of the contraction-free quantum state observation. 

Finally we provide the transmission function for the configuration superposition state.
Since the transmission function of a single configuration (i.e., A or B) $T^{A(B)}$ is written as 
\begin{eqnarray}
T^{A(B)} &=& \left|\left\langle 1 |{\rm \bf \Psi}^{A(B)} \right\rangle\right|^2       \nonumber   \\
&=&\left| (G^{R}_{0})_{1,1} (\sum_{j} c^{A(B)}_j \left\langle 1| W | \phi^{A(B)}_j \right\rangle )\right|^2,
\label{eq17}
\end{eqnarray}
the transmission function of the superposition state between the configurations A and B is 
\begin{eqnarray}
T^{A+B} &=& \left|\left\langle 1 |{\rm \bf \Psi}^{A+B} \right\rangle\right|^2       \nonumber   \\
&\simeq & \frac{1}{2} (T^{A}+T^{B}) .
\label{eq18}
\end{eqnarray}
In the last transformation, we omitted the contribution from the cross term of $\sqrt{T^{A}}\sqrt{T^{B}}$ because of the negligibly small number of the term when A or B (or both) corresponds to the destructive interference. Eq.~(\ref{eq18}) corresponds to a simple average between $T^{A}$ and $T^{B}$, and the result is originated from the identical weight of the configurations A and B in Eq.~(\ref{eq14}), which corresponds to $\theta = 45^\circ $ in the unitary operator shown in Fig. 4B.  We adopted Eq.~(\ref{eq18}) in the calculation of the transmission function for the configuration superposition shown in Fig. 3B.

\section{Conclusion}
In summary, we proposed a new system and theory for quantum computing using SMC, in which a single molecule is loosely sandwiched between electrodes. The striking features of this system are (i) an individual molecule in SMC can be regarded as a sequence of quantum-gates derived from the molecular orbital rule for quantum tunneling (i.e., an individual molecule in SMC corresponds to an individual quantum computer), (ii) the quantum-mechanical results as quantum tunneling states can be observed without contraction of the superposition states (i.e., contraction-free observation of quantum states), and (iii) quantum computing with SMC can be performed at room temperature. In this study, the adenine molecule was adopted as the single molecule in SMC, and the conductance data of the adenine-SMC was encoded onto quantum states, depending on the conductance. Single molecule identification was explained as an application of this quantum computing system, where a back-flow mode of quantum computing (i.e., the backward time propagation of quantum tunneling) with encoded quantum states based on observed conductance was introduced; existing quantum-gate quantum computers can be used for quantum computing with the back-flow mode.

\section{Acknowledgments}
\begin{acknowledgments}
The present work was supported in part by KAKENHI (the Grant-in-Aid for Challenging Exploratory Research No. 19K22044) to T.T. and by KAKENHI (Grant-in-Aid for Scientific Research(A) No. 19H00852) to M.T.
\end{acknowledgments}

\section{Appendix}
\appendix
\renewcommand{\thefigure}{A1}
\renewcommand{\thetable}{A1}

\section{A1. Molecular Hamiltonian}
In this section, we introduce the tight-binding Hamiltonian for both the adenine single molecule and the SMC configuration. The molecular structure of the adenine single molecule was optimized using the density functional method with B3LYP hybrid functional and 6-311G(d,p) basis set.\cite{G09} The calculated frontier orbitals (HOMO and LUMO) are shown in FIG. A1, from which the frontier orbitals are recognized as $\pi$-MOs. Thus, we constructed a tight-binding Hamiltonian in the $\pi$-orbital approximation for adenine, $\rm \bf H_{A}$. The Hamiltonian matrix (symmetric one) of adenine is written as
 \begin{eqnarray}
{\rm \bf H_{A}} &= &  \nonumber   \\
&  & \bordermatrix{
 &  1& 2 &   3 &  4  &5 & 6 & 7 & 8& 9& 10 \cr
1 &   0 &  &     &     &  &   &   &  &  &   &\cr
2 &   b & g &     &     &  &   &   &  &  &   &\cr
3 &   f & g &  a'   &     &  &   &   &  &  &   &\cr
4 &   d & c &  h   &  0   &  &   &   &  &  &   &\cr
5 &   c & b &  a   &  b  & a &   &   &  &  &   &\cr
6 &   0 & a &  0   &  a  & b &  a' &   &  &  &   &\cr
7 &   0 & 0 &  0   &  0  & c &   b & a'  &  &  &   &\cr
8 &   0 & 0 &  0   &  a  & 0 &   a & b  &  a' &  &   &\cr
9 &   0 & 0 &  a   &  b  & c &   0 & c  &  b& a &   &\cr
10 &   0 & 0 &  0   &  a  & 0 &   0 & 0  &  a& b &  e &\cr
}, \nonumber   \\
&  &
\label{eq1}
\end{eqnarray}
where the serial numbers, 1, 2, $\cdots$, 10 correspond to the atom indices of adenine (See Fig. A1A), and the parameters $a, a', b, c, d, e, f, g,$ and $h$ are respectively 0.1, -0.1, -1.0, -0.2, -0.3, -0.4, -1.1, -0.6, and -0.9, which were extracted  from the Fock matrix of the B3LYP/6-311G(d,p) level of calculation. In $\rm \bf H_{A}$, 4.07 eV is scaled to be 1, and the on-site elements (i.e., the diagonal elements) are shifted so as to be 0 of ${\rm \bf H_{A}}_{(1,1)}$. The frontier orbitals obtained by the diagonalization of $\rm \bf H^{A}$ are depicted in Fig. 2B. Note that the number of $\pi$-electrons of adenine is 12. 

For the adenine SMC, we have to determine the configurations of adenine coupled with gold electrodes. Since we know that the nitrogen atoms numbered by 3, 6, and 8 in Fig. A1A can make a contact with gold, we first optimized the positions of a single gold atom with respect to the nitrogen atoms. To make a minimum model for SMC configurations, we introduced two gold atoms; each gold atom is contacted with a nitrogen atom of adenine. Among the possible combinations for the contact positions of gold (i.e., Au-[N(3);N(6)]-Au, Au-[N(3);N(8)]-Au, and Au-[N(6);N(8)]-Au, where the notation of Au-[X($n$);X($m$)]-Au indicates that $n$- and $m$-th X atoms of adenine are used for the contact with gold), we found that gold-gold distance becomes the largest one when the configuration of Au-[N(3);N(6)]-Au is selected for the contact; the gold-gold distance is about 8 \AA. By fixing the gold-gold distance to be  8 \AA, we calculated the total energies for other contact configurations.  The calculated total energies are listed in Table A1, and the configurations of Au-[N(6);N(3)]-Au and Au-[N(6);N(1)]-Au are found to be plausible ones as the adenine SMC; the former and latter ones correspond to the constructive and destructive pathways, respectively. The matrix element of Hamiltonian for the off-diagonal (i.e., interaction) term between 6s(Au) and 2p$_{\pi}$(N/C) is determined to be 0.3 in accordance with the structures.  The transmission functions shown in Fig. 2A were calculated using the parameters.  

\begin{figure}
\includegraphics[scale=0.5]{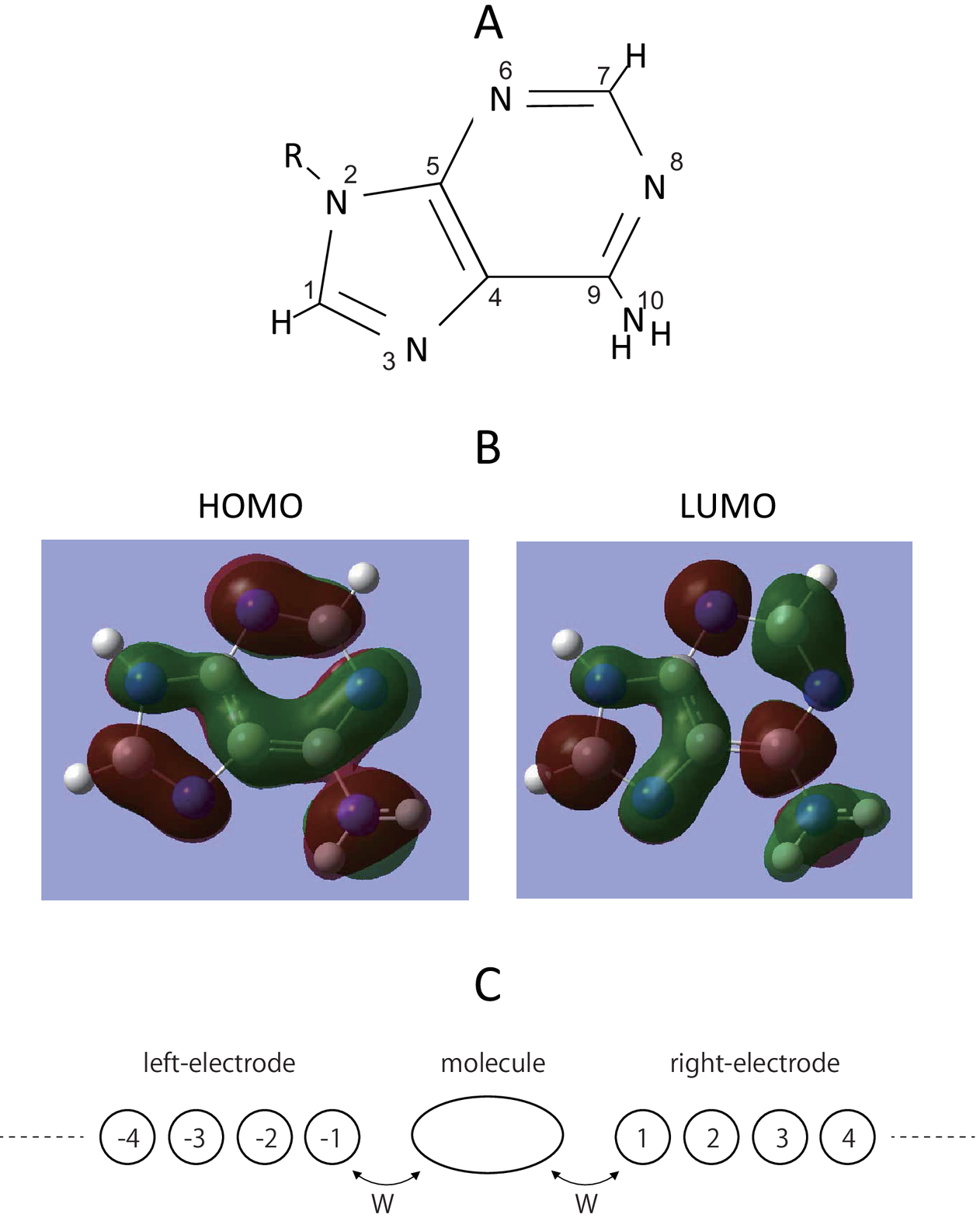} 
\caption{\label{molc} A: Atom indices of the adenine molecule. B: Calculated HOMO and LUMO with B3LYP/6-311G(d,p). C: Schematic of a single molecular confinement}
\end{figure}

The above procedure is enough to determine the matrix elements for a single molecule and its SMC, and in fact the theoretical foundation for the contraction-free quantun state encoding is not affected by the details of the parameters. However, we carried out one more additional step to obtain more realistic parameters especially for the electrode-molecule contact. Since we don't know the exact gap-width between electrodes, the procedure described in the previous paragraph cannot be a sufficient one because the gold-gold distance was determined in an artificial way.  In fact, the calculated conductance for the constructive pathway in Fig. 2 is clearly larger than the maximum conductance in experiment by a few orders of magnitude. In other words, an actual gap-width must be larger than the artificial one, 8 \AA. Thus, we have to evaluate the gap-width between electrodes in another way.  In this study, we adopted the follwing steps: i) the gap-width is gradually enlarged from 8 \AA, ii) we calculate the off-diagonal element between 6s(Au) and 2p$_{\pi}$(N/C) with the enlarged gap-width, iii) we calculate the conductance for the constructive pathway (i.e., Au-[N(6);N(3)]-Au) using the enlarged gap-width, and iv) we determine a gap-width that leads to the same order of the maximum conductance in experiment. The determined gap-width in the procedure is 8.5 \AA, and the  off-diagonal terms are 0.03 for 6s(Au)-2p$_{\pi}$(N(3,6) and C(1)). The transmission functions shown in Fig. 3 were obtained by using the parameters.

\begin{table}
\caption{\label{tab:tableS1} Calculated total energies of Au-adenine-Au SMC configurations with Au-Au distance of 8 \AA.}
\begin{ruledtabular}
\begin{tabular}{lcccr}
Configuration & Total energy (hartree) \\
\hline
Au-[N(6);N(3)]-Au & -738.1771 \\
Au-[N(6);C(1)]-Au & -738.1757 \\
Au-[N(6);N(10)]-Au & -738.0866 \\
Au-[N(3);C(7)]-Au & -738.1658 \\
Au-[N(3);N(8)]-Au & -738.1642 \\
Au-[N(8);C(1)]-Au & -738.1619 \\
Au-[N(8);N(2)]-Au & -738.1558 \\

% & - & ~4.53\footnote{Experimental data in Ref. \cite{rubin}} & - & -~~~~\\
\end{tabular}
\end{ruledtabular}
\end{table}

\section{A2. Green's function method for conductance}
In this section, we briefly introduce the non-equilibrium Green's function method for the conductance of a molecular-contact.\cite{datta} A schematic of a molecular contact in which the single molecule is sandwiched between the left- and right-electrodes is shown in Fig. A1C. As the first step, we consider a single molecule interacted with the right electrode only, and we apply the result of the one-electrode model to the two-electrode model. Since the system can be represented with a tight-binding Hamiltonian, the Green's function for the one-electrode (i.e., the right-electrode) system can be represented as 
 \begin{eqnarray}
{{\rm \bf G}(E)} &= & ((E+i0^{+}){\rm \bf I} - {\rm \bf H})^{-1}  \nonumber   \\
& = & \bordermatrix{
   &     &    &    \cr
   &   (E+i0^{+}){\rm \bf I} - {\rm \bf H_{M}} &  \vdots  &  - {\rm \bf   H}_{int}      \cr
   &   \cdots     \cdots \cdots        \cdots \cdots    &               &     \cdots \cdots  \cdots \cdots  \cdots    \cr
   &                - {\rm \bf H}_{int}^{+}                     &    \vdots &             (E+i0^{+}){\rm \bf I} - {\rm \bf H_{R}}     \cr
}^{-1},\nonumber   \\
& & \label{eq2}
 \end{eqnarray}
where  $\rm \bf H_{M}$ and $\rm \bf H_{R}$ are respectively the Hamiltonian matrices of the sandwiched single molecule and right-electrode, and  ${\rm \bf H}_{int}$ is the matrix representing the interactions between them (i.e., W in Fig. A1C). $\rm \bf I $ is the unit matrix and $0^{+}$ is an infinitesimally small number with the positive sign.  When the single molecule is adenine, $\rm \bf H_{A}$  can be used as $\rm \bf H_{M}$. Using Eq.~\ref{eq2}, we obtain Green's function of the single molecule part as

\begin{equation}
{{\rm \bf G}_{M}(E)} = (E{\rm \bf I} - {\rm \bf H_{M}} - {\rm \bf \Sigma}_{R} (E))^{-1} , 
\label{eq3}
\end{equation}
where the self-energy of the right electrode $ {\rm \bf \Sigma} _{R}(E) $ is represented as
\begin{equation}
 {\rm \bf \Sigma} _{R}(E) = {\rm \bf   H}_{int}  ((E+i0^{+}){\rm \bf I} - {\rm \bf H_{R}})^{-1} {\rm \bf   H}_{int}^{+}  .
\label{eq4}
\end{equation}
Since the term  $((E+i0^{+}){\rm \bf I} - {\rm \bf H_{R}})^{-1}$ is the Green's function of the electrode, we can easily calculate the Green's function of the electrode depending on the structure of electrodes (e.g., Fig. A1C is an example for the one-dimensional electrode, and the Green's function can be represented in an analytical equation \cite{emberly}), and we can straightforwardly obtain the self-energy. Now that we have recognized the influence by the electrode can be simply introduced via the self-energy as shown in Eq.~\ref{eq3}, we can apply the result of the one electrode model to the two electrodes as,
\begin{equation}
{{\rm \bf G}_{M}(E)} = (E{\rm \bf I} - {\rm \bf H_{A}} - {\rm \bf \Sigma}_{R} (E) - {\rm \bf \Sigma}_{L} (E))^{-1},
\label{eq5}
\end{equation}
where  $ {\rm \bf \Sigma} _{L}(E) $ is the self-energy of the left- (i.e., the second) electrode. Using the Green's function of molecule ${{\rm \bf G}_{M}(E)}$ and the  self-energy of electrodes, the transmission function $T$ can be represented as

\begin{eqnarray}
T(E) &= & {\rm Tr}[i\{{\rm \bf \Sigma}_{L}(E) -{\rm \bf \Sigma}^{+}_{L}(E)\} {\rm \bf G}_{M}(E) \nonumber   \\
& \times & ~~~~  i\{{\rm \bf \Sigma}_{R}(E) -{\rm \bf \Sigma}^{+}_{R}(E)\} {\rm \bf G}_{M}^{+}(E)] ,
\label{eq6}
\end{eqnarray}
where Tr[$\rm \bf A$] is the trace of matrix $\rm \bf A$.

\section{A3. Experimental conditions for electrical measurement in single molecule confinement}
We first describe the sample preparation. We purchased dAMP (2'-deoxyadenosine 5'-monophosphate from Sigma-Aldrich) to prepare 0.10-$\mu$M aqueous sample solutions without further purification. The prepared sample solutions including adenine nucleotide were inserted into the solution chamber of a nanofabricated mechanically controllable break junction (MCBJ) device,\cite{taniguchi3} which was produced using a nanofabrication technique.

A nanofabricated MCBJ was used for electrical measurements. The MCBJ substrate was bent, and the junction was mechanically broken to form a pair of gold nanoelectrodes. After reconnecting the gold junction, a constant DC bias voltage of 0.1 V was applied, and the substrate was gradually bent using a piezo-actuator. While breaking the junction, the junction conductance ($g$) was monitored using a picoammeter (Keithley 6487). A series of conductance jumps of the order of $g_0 = 2e^2/h$ (where e and h are the electron mass and Planck constant, respectively) were observed, and the final conductance was 1 $g_0$. Several seconds after the 1-$g_0$ state was obtained, the single gold atom contact naturally ruptured, resulting in a pair of gold electrodes. The gap width was controlled by tuning the piezo-voltage, and the electrode gap was set to an optimal value for the adenine nucleotide measurement. The current across the gap electrodes was recorded at 10 kHz using a custom-built transimpedance amplifier and a PXI-4081 digital multimeter (National Instruments) at a DC bias voltage of 0.1 V. Based on the observed current profiles, the gap width was maintained at an optimal value during the entire run as follows: every 0.5 s, the baseline current was defined as the mode value of the observed current-time profile of the 5,000 data points. The baseline current in the observed current profile represents the tunnel current of the gap at that time.

%The tunnel current (I) can be expressed using the following equation:

%\begin{equation}
%I = exp(-const. \times d \sqrt{\phi})
%\label{eq19}
%\end{equation}
%where $d$ is the gap width and $\phi$ is the work function of the gap electrode.
%From this equation, it is observed that the tunnel current is exponentially proportional to the gap width. To sustain the set gap width, the baseline current was maintained by adjusting the piezo-voltage during the entire run.

 % \left\langle\varphi|\psi\right\rangle

%\begin{acknowledgments}
%\end{acknowledgments}


\end{document}